# DEMO Hot Cell and Ex-Vessel Remote Handling

Justin Thomas[a], Antony Loving[a], Christian Bachmann[b], Jon Harman[b]

[a] *Euratom/CCFE Fusion Association, Culham Science Centre, Abingdon, OX14 3DB, UK*

[b] *EFDA, PPP&T, Boltzmannstraße 2, 85748 Garching, Germany*

In Europe the work on the specification and design of a Demonstration Power Plant (DEMO) is being carried out by EFDA in the Power Plant Physics and Technology (PPP&T) programme. DEMO will take fusion from experimental research into showing the potential for commercial power generation.

During the fusion reaction, components in the tokamak become highly activated. The estimated dose rate levels after shutdown (zero decay time) due to 60 dpa accumulation in steel (blanket) and 30 dpa (divertor) are 13.1-17.4 kGy/h (blanket); 8.8-11.6 kGy/h (divertor)[1], much higher than those to be encountered at ITER.

Upon removal from the tokamak, components would be transported to the Hot Cell facility with attention to minimizing the spread of activated dust and tritium contamination. It is proposed to use a sealed cask of ~20 tonnes, running on air castors with 50% lifting capability redundancy. Due to the number and complexity of the routes taken by this transporter it would have to be an un-tethered semi-autonomous system. This poses some technical challenges, including providing sufficient battery capacity, reliable guidance and a fail safe un-tethered control system. The mass of the components being moved is assumed here to range from a few tonnes to in excess of one hundred tonnes.

Before the removed in-vessel components can be processed in the Hot Cell, they would require a period of cooling, approximately 2.5yrs, to allow dose rate and decay heating to reduce. This reduces the decay heating level to ~1-1.5 kW/m³ and a contact dose rate to ~250Sv/hr, which is more suited to dexterous man-in-the-loop Remote Handling (RH). During maintenance, many components would require replacement or refurbishment. The Hot Cell facility would have to provide all the associated RH functions and operate fully remotely. With no human intervention, an accordingly robust RH recovery system would be required.

This paper describes the first steps being taken towards the design of the DEMO Hot Cell. It will show a comparison of the current DEMO in-vessel maintenance concepts from a Hot Cell perspective, describe a proposed ex-vessel transport system, and summarize the facilities that have been identified as required within the Hot Cell, examine current RH technology and discuss the identified critical development issues.

## 1. Introduction

DEMO is to be the next stage after ITER. It will be used to demonstrate that fusion can be used to commercially produce electricity. The design of this facility is being started before ITER has finished operating so there is some continuity between projects. All of the technologies and facilities for a commercial power station will be brought together. During the plasma reaction the internal component of the tokamak become highly activated. The resultant activation makes the in-vessel components very difficult to handle. The dose rate and decay heating reduce with time, but the levels could take approximately 2.5yrs to fall to a level that is more suited to dexterous man-in-the-loop remote handling.

## 2. DEMO Maintenance Concepts

During the operational life of DEMO many components will require maintenance or replacement. Maintenance of in-vessel components will have to be done in a hot cell using fully remote handling techniques, this is due to their level of activation. Manned interventions might not be possible when vessel components are present in the hot cell. The remote handling systems will require an accordingly robust recovery system. Where possible, redundant systems would be used so the equipment can recover itself to the RHE maintenance area.

A number of maintenance concepts have been proposed for the DEMO machine. Some of these are described below.

**Multi Module Segment (MMS)** [2]

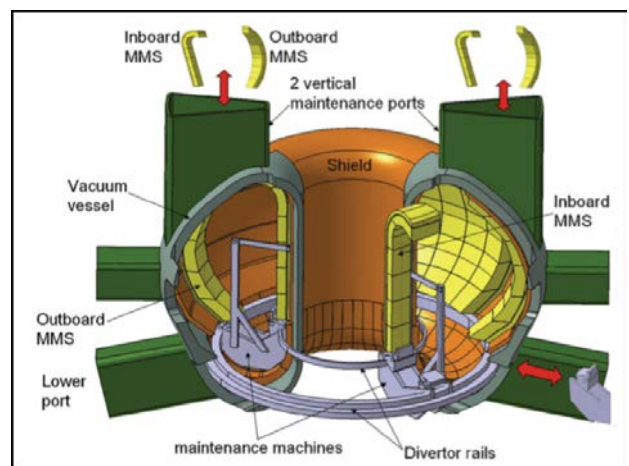

**Figure 1 Multi Module Segment** [2]

For this concept the vessel walls would be divided in to segments with a separate divertor. The segments would be split between the inner and outer walls, totalling 50 to 80. The divertor is comprised of 30 cassettes. The segments would be removed through vertical ports in the top of the tokamak and the divertor through horizontal ports at the bottom. Multiple segments would be removed through a port allowing space in between the ports for the coils.

Because of their smaller size the segments can be packed more densely when in storage. This would reduce the size of the required storage facility.

**Large Sector (Vertical)** [3]

The vacuum vessel would be divided in to 36 sectors, each of which is a full 'D' containing the inner wall, outer wall and the divertor. Each sector is removed through a port in the top of the machine. The basic vacuum vessel structure remains in place during removal. Each sector is estimated to have a mass of 130 tonnes although this could increase to ensure the sector is strong enough to support its own weight.

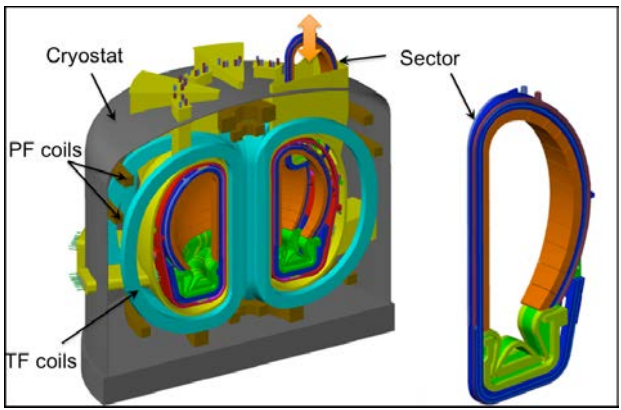

**Figure 2 Large Sector (Vertical)**

**Large Sector (Horizontal)** [3]

A full sector, one of 12 slices through the machine, would be removed horizontally through a port in the cryostat. The sector includes in-vessel and ex-vessel structures which are handled as one large component. It is estimated that each sector would have a mass of 720 tonnes. Moving such large sectors would require much more space in the reactor building.

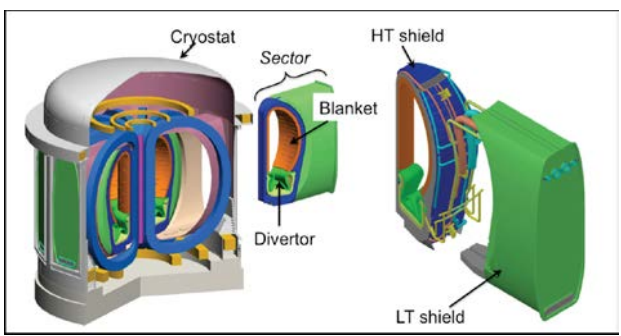

**Figure 3 Large Sector (Horizontal)**

The size and position of the ports that the sectors are removed through compromise the position of the poloidal field coils. An innovative solution is required to allow the super cooled coils to be moved during maintenance to allow optimum positioning when operating.

Both large sector concepts are not ideal for the hot cell. The space required for storage would be similar to the area occupied by the reactor, the sectors have a large volume of air that can not be utilized for storage.

The use of large sectors would make the removal of the in-vessel components simpler. But all of the problems with high dose rates and decay heating would become a problem for the hot cell. Clearly these problems can be eliminated so the effort put in to this early work will guide the decision on where it is best dealt with.

## 3. Ex-Vessel Transport System

The ex-vessel transport could be similar to the system being used at ITER. This system is based on casks. The casks are mounted on a transporter.

The proposed system could have the casks and transporters separate and interchangeable. This would give greater flexibility to the system. It could reduce downtime due to maintenance and potentially reduce costs. Fewer transporters would be required as the casks would outnumber them.

The transporters would have to work in many areas around the reactor and hot cell buildings. Because of this a number of options were discounted. Due to the number of interlinking routes that can be used and parallel operation, the transporter would have to be battery powered and semi-autonomous.

The mass of the transporter and its load could be carried on air castors. Pre-loaded wheels would provide the steering and motive power, but would not carry any of the loads.

The casks have to perform a number of critical functions and therefore can not be considered to be simple passive containers. Depending on the components being carried cooling would need to be provided to offset the decay heating of the component. This system would need to be able to function in the event of a breakdown in the transport system. The cask would not need to provide shielding for the workforce as timed movements and the buildings would provide this. However the transporter control and safety systems would need to be protected from the highly activated in-vessel components that would be moved using this system.

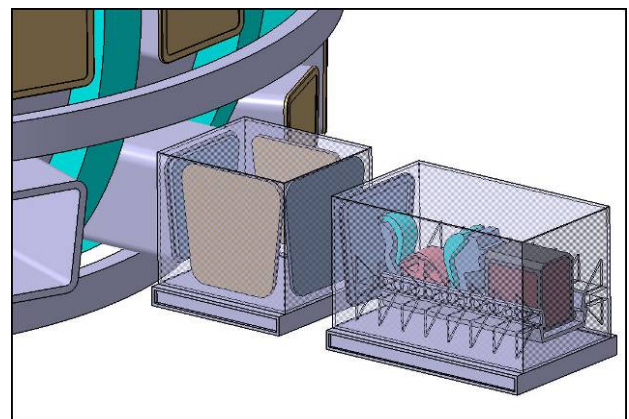

**Figure 4 Door and Cask Systems**

A number of different transport methods were considered with air castors having the most favourable characteristics. Although a wheeled system would be more than capable of carrying the predicted loads, they also come with some key constraints. A railway system would require many switches in the track. This adds complexity and potential failure points to a system that has to operate in areas that people would not be able to enter. The turning radius of a vehicle on rails is quite large. Turntables would have to be used to reduce this to fit within the space constraints. This would add further complexity and potential reliability issues. Multi wheel transporters have the ability to turn on their axis, so confined spaces are not an issue. However the steering and suspension systems are complex. Each wheel is independently sprung and steered. Current available systems use hydraulics for these functions, not ideal for the anticipated operating environment. Many tires, potentially 50, would create a lot of dust and debris in areas that are not easy to maintain. Another consideration with wheeled systems is the minimum height that can be achieved.

Due to the complexity of the routes and the number of transporters in use at the same time a wired power supply would not be possible, power would come from batteries. The batteries would require sufficient capacity to power the control systems and motive systems for a round trip from the charging station. Ideally charging would be carried out in one area away from the reactor hall.

The transporter would guide itself. It would follow cables embedded in the floor. In the event of loss of guidance the control system would immediately go to safe state. Manual remote control should be an option in the event of a loss of the guidance system or signal. Additional safety systems should be implemented to recognize deviations from the intended path. The space available to operate would be limited, so any deviation from the intended path would soon result in a collision.

For ports that the cask docks to multiple times, a separate door system might be beneficial [2]. It would take the complexity of the door out of the cask. This reduces the number of door systems that need to be purchased. The door would be docked to the vessel port before the first cask is attached and remain in place until all of the activities at that port are complete. This would also release the space in the cask required to carry the equipment associated with the door. This results in the cask having a larger capacity or the being a smaller size.

### 4. Hot Cell Facilities

For DEMO the Hot Cell would be a group of facilities that perform many functions. It is not just a room where active components are worked on. The capacity of the Hot Cell would have to be large enough to receive and process components at a rate equal to or greater than the rate at which they are removed from the machine. This flow of components would come from both in-vessel and ex-vessel. This capacity would have to be maintained for the duration of a shutdown regardless of planned and un-planned maintenance operations.

A number of key functions have been identified for the Hot Cell facility. Due to the scale of the Hot Cell facility it might be concluded that it is more cost effective to share these facilities between multiple power plants.

The Hot Cell Facility is likely to be larger than the reactor building, especially if a large sector maintenance concept is adopted. Figure 5 shows the scale of the hot cell building next to the reactor building. The image shows one level of the facility containing large component maintenance, initial wet storage ponds and post maintenance dry storage. Further levels would be required to house all of the different facilities required.

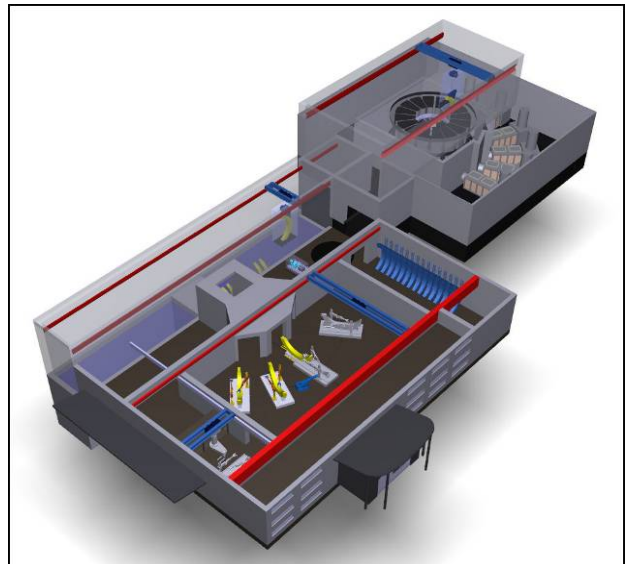

**Figure 5 Hot Cell and Reactor Buildings** [2]

Many functions for the DEMO hot cell were identified during the initial concept discussions. Below is a list of the facilities that would be required in power plant hot cell. The facilities required would be storage, maintenance, transfer, cleaning, inspection and waste processing. Figure 6 shows a concept of how these facilities could be laid out.

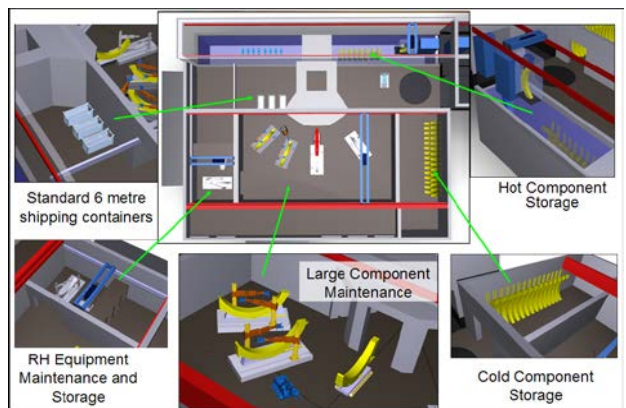

**Figure 6 Hot Cell Facilities Concept** [2]

### 5. Current Remote Handling Technologies

**Mechanical through the wall manipulators.**

The Mechanical Master Slave Manipulator is also known as a Through-the-Wall Manipulator. This system is widely used for the processing and handling of highly contaminated and radioactive items. This is a well proven and developed technology, widely used in a variety of industries and has been for many years.

The manipulator does not rely on electronics for its operation. Viewing of the work is done directly by the operators through a shielded window.

The working area of the manipulator is restricted to the maximum movement of the arms. This could be improved by moving the item under manipulator. Also telescopic arms can be used to increase the work area. This does reduce the load capacity of the arm though.

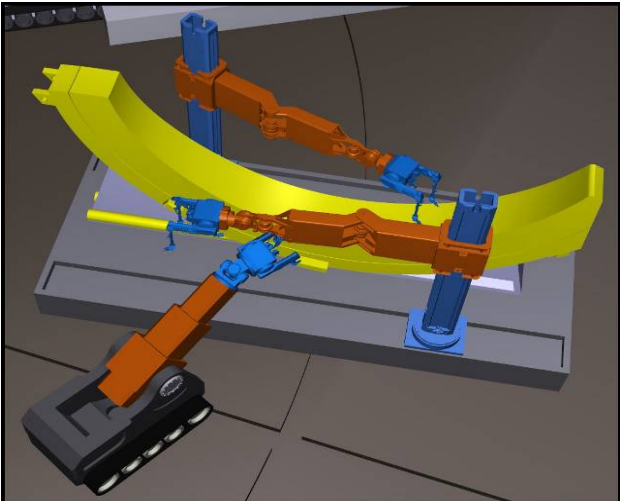

**Figure 7 Large Component Palletized Maintenance [2]**

**Electrical Force Reflecting Servo Manipulators.**

The master and slave arms are connected electrically. They are powered by electric motors and have a digital control system. The operator inputs to the master are interpreted by the control system and duplicated on the slave. The operator is provided with force feedback, similar to that of a mechanical system. A standard computer is used by the operator to access and activate enhanced functions.

The control system can provide the operator with many useful features that can improve performance and reduce fatigue. A variable force ratio can be applied to reduce load moved by the operator. Weight compensation can be used to completely remove the load that the operator feels. Moves and positions can be accurately saved and replicated. Accurate torque values can be derived from the system up to the capacity of the manipulator. The control system is programmable allowing new features to be added as required.

The slave manipulator can be mounted on a transporter to increase the working area to match the requirements of each task. The transporter can be attached to the building or mounted on a vehicle. This can make this type of system reliant on video feedback.

The operator can be located almost anywhere on site as viewing id done using video. Viewing through windows is difficult because of the large working area.

The manipulator movements can be displayed in a Virtual Reality (VR) environment in real-time using outputs from the control system. VR provides valuable visual feedback that supplements camera views when operating in complicated and confined environments.

A flexible and adaptable remote handle system would be required to cope with the large variation in size and complexity of components that would be maintained.

**Automated Maintenance Systems.**

Fully automated maintenance systems as required by DEMO are not readily available. The nearest technology is production line machines. These machines are designed to do specific repetitive tasks at high speed. Fully automated systems can only operate within parameters that are known. Unforeseen problems such as melting or distortion of a component could cause the system to stop or misinterpret the environment.

An automated system could be suited to measurement and inspection tasks. In particular non-contact inspection methods using video or laser technologies. An automated system would be able to conduct these surveys faster and with greater repeatability than a hand operated system.

## 6. Conclusions

Making accurate estimations of dose rates and decay heating is difficult at the moment due to the lack of detail in the in-vessel component designs. These estimates will be refined as the composition and design of components evolves.

The limited radiation hardness of dextrous remote handling systems could be a factor when designing the hot cell. Video cameras in particular systems do not currently offer enough tolerance to radiation.

With such a reliance on fully remote operation, the RH systems would require significant redundancy. Where this is not possible, recovery systems and detailed procedures would need to be developed and validated.

The construction of a physical mock-up and test facility in the early design stage should be considered. Using such a facility to validate component and equipment designs before going in to full production could save time and resources.

## References

[1] WP11-DAS-RH-07, Definition of Radiation Map in DEMO. U. Fischer, Karlsruhe Institute of Technology (KIT), Germany, J. Sanz, J.P Catalán, R. Juárez, Universidad Nacional de Educación a Distancia (UNED)/CIEMAT, Madrid, Spain

[2] WP11-DAS-RH-05 Identification of DEMO Ex-Vessel and Hot Cell Maintenance Operations. Justin Thomas Culham Centre for Fusion Energy (CCFE)

[3] H. Utoh et.al, Conceptual study of vertical sector transport maintenance for DEMO fusion reactor, JAEA

This work is based on current conceptual assumptions, not an official DEMO design.

This work was funded by the RCUK Energy programme and EURATOM.

The work was carried out within the framework of EFDA